\documentstyle[12pt]{article}
\begin{document}
\title{{\bf ON THE VARIATIONAL}
{\bf CHARACTERISATION OF}\\
{\bf GENERALIZED JACOBI EQUATIONS}\footnote{This article was originally published in {\it Differential Geometry and Applications}, Proc. Conf., Aug. 28 -- Sept. 1, {\tt 1995}, Brno, Czech Republic (Masaryk University, Brno, {\tt 1996}), pp. 353--372. Since it is hardly available we post it here for the interested audience.}}

\author{{\bf Biagio Casciaro}\footnote{\tt casciaro@pascal.uniba.it}\\
\\
{\it Dipartimento di Matematica}\\
{\it Universit\`a di Bari}\\
{\it via Orabona 5, 75125 Bari, Italy}\\
\\
{\bf Mauro Francaviglia}\footnote{\tt francaviglia@dm.unito.it}\\
\\
{\it Istituto di Fisica Matematica ``J.--Louis Lagrange''}\\
{\it Universit\`a di Torino}\\
{\it via Carlo Alberto 10, 10123 Torino, Italy}\\
\\
and\\
\\
{\bf Victor Tapia}\footnote{\tt tapiens@gmail.com}\\
\\
{\it Departamento de Matem\'aticas}\\
{\it Universidad Nacional de Colombia}\\
{\it Bogot\'a, Colombia}}

\maketitle

\newpage
\begin{abstract}

We study higher--order variational derivatives of a generic second--order Lagran\-gian ${\cal L}={\cal L}(x,\phi,\partial\phi,\partial^2\phi)$ and in this context we discuss the Jacobi equation ensuing from the second variation of the action. We exhibit the different integrations by parts which may be performed to obtain the Jacobi equation and we show that there is a particular integration by parts which is invariant. We introduce two new Lagrangians, ${\cal L}_1$ and ${\cal L}_2$, associated to the first and second--order deformations of the original Lagrangian ${\cal L}_0$ respectively; they are in fact the first elements of a whole hierarchy of Lagrangians derived from ${\cal L}_0$. In terms of these Lagrangians we are able to establish  simple relations between the variational derivatives of different orders of a given Lagrangian. We then show that the Jacobi equations of ${\cal L}_0$ may be obtained as variational equations, so that the Euler--Lagrange and the Jacobi equations are obtained from a single variational principle based on the first--order variation ${\cal L}_1$ of the Lagrangian. We can furthermore introduce an associated energy--momentum tensor ${{\cal H}^\mu}_\nu$ which turns out to be a conserved quantity if ${\cal L}_0$ is independent of space--time variables.

\end{abstract}

\setcounter{section}{-1}
\section{Introduction}

As is well known, the second variation of an action functional governs the behaviour of the action itself in the neighbourhood of critical sections. In particular, the Hessian of the Lagrangian defines a quadratic form whose sign properties allow to distinguish between minima, maxima and degenerate critical sections (see, {\it e.g.}, \cite{01}). It is also well known that in the case of geodesics in a Riemannian manifold those fields which govern the transition from geodesics to geodesics, ({\it i.e.}, those vectorfields which make the second variation to vanish identically modulo boundary terms) are {\it Jacobi fields} \cite{02} and they are solutions of a second--order differential equation, namely the {\it Jacobi equation} (of geodesics).

The notion of Jacobi equation as an outcome of the second variation is in fact fairly more general than this and general formulae for the second variation and generalised Jacobi equations along critical sections have been already considered in the Calculus of Variations from a ``structural viewpoint'' (see, {\it e.g.}, the review of results contained in \cite{03}). To our knowledge, however, in most of the current literature on the subject the second variation of functionals has been considered in a direct way and without resorting to general expressions, while integration by parts to reduce it to more suitable forms have been usually performed by {\it ad hoc} procedures, in spite of the fact that fairly general such formulae can be worked out (see below for a comment on a paper by Taub \cite{04}, which is appropriate to mention here but is better to discuss later). Because of these facts, we have reached the conclusion that the theory of second variations is worthy of being revisited, also in view of a number of applications which shall be mentioned later and will form the subject of forthcoming investigations.

In a previous paper \cite{05}, working in the general and global framework of first--order variational principles in fibered manifolds and their jet--prolongations, the notion of {\it generalised Jacobi equations} was developed and discussed for first--order Lagrangians. As is well known, the classical Jacobi equation for geodesics of a Riemannian manifold defines in fact the Riemannian curvature tensor of $g$; because of this we can say that the second variaton $\delta^2{\cal S}$ and the generalised Jacobi equations define the ``curvature'' of any given variational principle. In the generic case, of course, this notion has very little to say, in a continuation \cite{06} of that work it was however shown that this general concept of ``curvature'' takes a particularly significant form in the case of generalised harmonic Lagrangians, giving rise to suitable ``curvature tensors'' which satisfy suitable ``generalised Bianchi identities''. Applications to second variations of relativistic Lagrangians ({\it i.e.}, Lagrangians depending on the full curvature of a Riemannian metric) are being considered in \cite{07} and \cite{08}.

For the sake of completeness, we should mention that a number of recent and less recent papers (see, {\it e.g.}, \cite{09}, \cite{10} and references quoted therein) have attacked the problem of ``generalised Jacobi equations'' and ``curvature'' for arbitrary second--order differential equations, from the general viewpoint of dynamical systems on a tangent bundle. These interesting contributions have a rather different nature from ours, since they refer to fairly general dynamical structures while the results we are going to discuss stem directly from the richer structure of Lagrangian systems on generic fiber bundles. Nevertheless, it would be interesting to compare the two viewpoints, also to check how much of our direct, simpler and straightforward results might be recovered from a suitable application of the more complicated structures discussed in \cite{10} (and references quoted therein) in a different and complementary framework. We hope to address this problem in future investigations.

In this paper we shall consider some further results related to the generalised Jacobi equation. For the sake of simplicity and for mere notational convenience we will establish our results only for second--order Lagrangians in field theory, although an analogous scheme can be worked out for higher--orders without bringing any really new insight into the problem.

The first part of this paper (Section 1) is devoted to briefly recall the situation for first--order theory, as already discussed in \cite{11} in the intrinsic language of differential geometry of jet--bundles. In Section 2 we consider the second variation $\delta^2{\cal S}$ of the action ${\cal S}$, defined by a Lagrangian depending at most on second--order derivatives of the fields. We then show how a number of different integrations by parts allow to recast the Hessian in more suitable forms, which contain the Euler--Lagrange equations and define some ordinary differential equations of the second--order which are the {\it generalised Jacobi equations}. These are in fact the equations which define along critical curves those vectorfields in the configuration space which make $\delta^2{\cal S}$ to vanish identically (modulo boundary conditions).

In Section 3 we introduce then a series of relevant relations between the Euler--Lagrange equation for a given Lagrangian ${\cal L}_0$, namely

\begin{equation}
{{\delta{\cal L}_0}\over{\delta\phi^A}}\equiv
{{\partial{\cal L}_0}\over{\partial\phi^A}}
-{d\over{dx^\mu}}\left({{\partial{\cal L}_0}\over{\partial{\phi^A}_\mu}}\right)
+{{d^2}\over{dx^\mu dx^\mu}}\left({{\partial{\cal L}_0}\over{\partial{\phi^A}_{\mu\nu}}}\right)=0\,,
\label{0.1}
\end{equation}

\noindent and the ``Jacobi equation'' for the same Lagrangian ${\cal L}_0$, namely

\begin{eqnarray}
J_A\equiv
{{\partial^2{\cal L}_0}\over{\partial\phi^A\partial\phi^B}}\,\eta^B
+{{\partial^2{\cal L}_0}\over{\partial\phi^A\partial{\phi^B}_\lambda}}\,{\eta^B}_\lambda
+{{\partial^2{\cal L}_0}\over{\partial\phi^A\partial{\phi^B}_{\lambda\rho}}}\,{\eta^B}_{\lambda\rho}&&\nonumber\\
-{d\over{dx^\mu}}\left(
{{\partial^2{\cal L}_0}\over{\partial\phi^A_\mu\partial\phi^B}}\,\eta^B
+{{\partial^2{\cal L}_0}\over{\partial\phi^A_\mu\partial{\phi^B}_\lambda}}\,{\eta^B}_\lambda
+{{\partial^2{\cal L}_0}\over{\partial\phi^A_\mu\partial{\phi^B}_{\lambda\rho}}}\,{\eta^B}_{\lambda\rho}
\right)&&\nonumber\\
+{{d^2}\over{dx^\mu dx^\nu}}\left(
{{\partial^2{\cal L}_0}\over{\partial\phi^A_{\mu\nu}\partial\phi^B}}\,\eta^B
+{{\partial^2{\cal L}_0}\over{\partial\phi^A_{\mu\nu}\partial{\phi^B}_\lambda}}\,{\eta^B}_\lambda
+{{\partial^2{\cal L}_0}\over{\partial\phi^A_{\mu\nu}\partial{\phi^B}_{\lambda\rho}}}\,{\eta^B}_{\lambda\rho}
\right)&=&0\,.
\label{0.2}
\end{eqnarray}

\noindent Starting from the Lagrangian ${\cal L}_0$ we shall in fact define new Lagrangians ${\cal L}_1$ and ${\cal L}_2$ associated respectively to the first--order deformation of ${\cal L}_0$ and to the second--order deformation of ${\cal L}_0$, with the remarkable property that the Euler--lagrange equations (\ref{0.1}) and the Jacobi equations (\ref{0.2}) of the original Lagrangian ${\cal L}_0$ can be rewritten together in terms of ${\cal L}_1$ and ${\cal L}_2$ as follows:

\begin{eqnarray}
{{\delta{\cal L}_0}\over{\delta\phi^A}}={{\delta{\cal L}_1}\over{\delta\eta^A}}&=&0\,,\\
{{\delta{\cal L}_2}\over{\delta\eta^A}}={{\delta{\cal L}_1}\over{\delta\phi^A}}&=&0\,.
\label{0.3}
\end{eqnarray}

\noindent Therefore, we can consider the above relations as coming from a single variational principle based on the new Lagrangian ${\cal L}_1$  where, however, $\phi$'s and $\eta$'s are considered as independent variables. This result finds important applications to Riemannian Geometry \cite{11}.

We can thence introduce momenta canonically conjugated to $\phi$'s and $\eta$'s by means of standard prescriptions (see Section 3 below) as well as a canonical energy--momentum tensor ${{\cal H}^\mu}_\nu$, which is conserved if ${\cal L}_0$ does not depend explicitly on ``space--time variables'' $x^\lambda$. Therefore, the allowed deformations are selected by the first--order condition

\begin{equation}
{{d{{\cal H}^\mu}_\nu}\over{dx^\mu}}=0\,,
\label{0.4}
\end{equation}

\noindent which replaces the second--order condition given by the Jacobi equation (\ref{0.2}).

We are now in position to mention in a greater detail the paper by Taub \cite{04} we already quoted above. In that paper, which is explicitly devoted to an application of second--order variations to relativistic fluid--dynamics, a general formula for the Jacobi equation is explicitly mentioned and it is claimed that a ``well--known result'' states that this equation is in fact the variational derivative of the Hessian ${\cal L}_2$ (see equation (\ref{0.3}) above. Unfortunately, no reference is given to the source of this result; we stress, however, that the viewpoint we shall discuss in this paper is fairly different and more general. In fact, we shall not see Jacobi equations alone as (partial with respect to $\eta$) Lagrange equations of the Hessian ${\cal L}_2$ (as in \cite{04}), but rather the system formed by Jacobi equations and the original Lagrange equations of ${\cal L}_0$ as variational equations of the first--order deformed Lagrangian ${\cal L}_1$.

Section 4 is finally devoted to some applications and examples.

For the sake of simplicity most of this paper will be written in the language of fundamental calculus in ${\bf R}^n$, although all the results presented here can be expressed and derived in the intrinsic language of fiber bundles.

\setcounter{equation}{0}
\section{The Lagrangian Characterization of Generalized Jacobi Equation in the First--Order Case}

Let us first recall some basic notion from the Calculus of Variations on fibered manifolds. Let $(B,M,\pi)$ be a fibered manifold and ${\cal L}_0:J^1\pi\rightarrow\Lambda^n(T^*M)$ be a fisrt--order Lagrangian (density) on $B$. Here $J^1\pi$ denotes the first--jet prolongation of $\pi$ and $\Lambda^n(T^*M)$ is the bundle of $n$--forms of $M$, $n$ being the dimension of the base manifold $M$. Locally

\begin{equation}
{\cal L}_0=L(x^\lambda,\,\phi^A,\,{\phi^A}_\lambda)\,ds\,,
\label{1.1}
\end{equation}

\noindent where $(J^1U;\,x^\lambda,\,\phi^A,\,{\phi^A}_\lambda)$ is any natural chart in $J^1\pi$ and $ds$ is the local volume of $(U;\,x^\lambda)$.

The action of ${\cal L}_0$ is defined by

\begin{equation}
{\cal S}=\int_\Omega\,(j^1\sigma)^*\,{\cal L}_0\,,
\label{1.2}
\end{equation}

\noindent where $\Omega$ is any compact domain in $M$ with regular boundary $\partial\Omega$ and $\sigma\in\Gamma(\pi)$ is any local section (defined in an open subset containing $\Omega$). One defines then the ``first variation'' $\delta{\cal S}$ of ${\cal S}$ by considering homotopic variations $\eta\equiv\delta\sigma\in\chi\nu(\pi)$ with fixed values at the boundary $\partial\Omega$, $\chi\nu(\pi)$ being the space of vertical vectorfields of $\pi$. The {\it critical sections} are those sections along which $\delta{\cal S}$ vanishes for any $\eta\in\chi\nu(\pi)$ with fixed values at the boundary. They are characterised by the equation:

\begin{equation}
(j^2\sigma)^*[e({\cal L}_0)]=0\,,
\label{1.3}
\end{equation}

\noindent which is the {\it Euler--Lagrange equation}. Here $[e({\cal L}_0)]$ is {\it Euler--Lagrange morphism}, a global bundle morphism $e({\cal L}_0):J^2\pi\rightarrow\Lambda^m(T^*M)\otimes V^*\pi$, where $V^*\pi$ is the dual bundle of the vertical bundle $V\pi$, locally defined by:

\begin{equation}
e({\cal L}_0)=e_A({\cal L}_0)\,ds\,\otimes\,\,d\phi^A\,.
\label{1.4}
\end{equation}

The first variation of ${\cal L}_0$ is in fact globally defined through a further global bundle morphism $f({\cal L}_0):J^1\pi\rightarrow\Lambda^{m-1}(T^*M)\otimes V^*\pi$, locally expressed by:

\begin{equation}
(j^1\sigma)^*f({\cal L}_0)=(p_A^\mu({\cal L}_0)\,\circ\,j^1\sigma)\,ds_\mu\,\otimes\,d\phi^A\,,
\label{1.5}
\end{equation}

\noindent where $ds_\mu$ is the surface element of $\partial\Omega$, defined so that $ds_\mu\wedge dx^\mu=(-1)^\mu ds$.

The following holds for $T{\cal L}_0$:

\begin{equation}
(j^1\sigma)^*[T({\cal L}_0)]=(j^2\sigma)^*\left<e({\cal L}_0)|\eta\right>+(j^1\sigma)^*d\left<
f({\cal L}_0)|\eta\right>\,,
\label{1.6}
\end{equation}

\noindent for any local section $\sigma$ and any vertical vectorfield $\eta$ which projects onto $\sigma$. Here and in the sequel $\left<|\right>$ denotes standard dualitiy between forms and vectorfields. Equation (\ref{1.6}) is the {\it global first variation formula} of ${\cal L}_0$. The integral counterpart of equation (\ref{1.6}) is thence the following

\begin{equation}
\delta{\cal S}=\int_\Omega\,\left<e({\cal L}_0)|\eta\right>+\int_{\partial\Omega}\,\left<f({\cal L}_0)|\eta\right>\,,
\label{1.7}
\end{equation}

\noindent from which the Euler--Lagrange equations (\ref{1.3}) follow after imposing the appropriate boundary condition $\eta|_{\partial\Omega}=0$.

In order to study the stabi;lity properties of critical sections, {\it i.e.} of the solutions of the Euler--Lagrange equation (\ref{1.3}), one considers next the variation of the action under second--order deformations of $\sigma$. The second variation of ${\cal S}$ is then given by (see, {\it e.g.}, \cite{05}):

\begin{equation}
\delta^2{\cal S}={1\over2}\,\left[\int_\Omega\,\left<e({\cal L}_0)|\rho\right>+\int_\Omega\,{\rm Hess}_{{\cal L}_0}(j^1\eta)+\int_{\partial\Omega}\,\left<f({\cal L}_0)|\rho\right>\right]\,,
\label{1.8}
\end{equation}

\noindent where $\rho=\delta^2\sigma$ denotes the second variation of $\sigma$ and the $n$--form ${\rm Hess}_{{\cal L}_0}(j^1\pi)$ is the {\it Hessian} of ${\cal L}_0$. Locally, equation (\ref{1.8}) reads as follows:

\begin{eqnarray}
\delta^2{\cal S}&=&{1\over2}\,\left[\int_\Omega\,\left[{{\partial{\cal L}_0}\over{\partial\phi^A
}}\,\rho^A+{{\partial{\cal L}_0}\over{\partial{\phi^A}_\mu}}\,{\rho^A}_\mu\right]\,ds\right.
\nonumber\\
&&\left.+\int_\Omega\,\left[{{\partial^2{\cal L}_0}\over{\partial\phi^A\partial\phi^B}}\,\eta^A
\,\eta^B+2\,{{\partial^2{\cal L}_0}\over{\partial\phi^A\partial{\phi^B}_\mu}}\,\eta^A\,{\eta^B}
_\mu+{{\partial^2{\cal L}_0}\over{\partial{\phi^A}_\mu\partial{\phi^B}_\nu}}\,{\eta^A}_\mu\,
{\eta^B}_\nu\right]\,ds\right]\,.
\label{1.9}
\end{eqnarray}

A second order equation for $\eta$ along critical sections, the {\it (generalised) Jacobi equation}, can be obtained from (\ref{1.8}) by suitable integrations by parts on ${\rm Hess}$. In fact, as discussed {\it e.g.} in \cite{05}, equation (\ref{1.8}) can be conveniently rewritten as follows

\begin{equation}
\delta^2{\cal S}={1\over2}\,\left[\int_\Omega\,\left<e({\cal L}_0)|\rho\right>+\int_\Omega\,
\left<{\rm Jac}_{{\cal L}_0}(j^2\eta)|\eta\right>+\int_{\partial\Omega}\,{\hat f}({\cal L}_0)
(\eta,\,\rho)\right]\,,
\label{1.10}
\end{equation}

\noindent where ${\hat f}({\cal L}_0)(\eta,\rho)$ is a new boundary term depending both on $\eta$ and $\rho$ and the term ${\rm Jac}_{{\cal L}_0}(j^2\eta)$ is locally given by

$${\rm Jac}_{{\cal L}_0}(j^2\eta)=J_A(j^2\eta)\,d\phi^A\,,$$

\noindent with

\begin{eqnarray}
J_A(j^2\eta)&=&{{\partial{\cal L}_0}\over{\partial\phi^A\partial\phi^B}}\,\eta^B
+{{\partial{\cal L}_0}\over{\partial\phi^A\partial{\phi^B}_\mu}}\,{\eta^B}_\mu\nonumber\\
&&-{d\over{dx^\mu}}\left({{\partial{\cal L}_0}\over{\partial{\phi^A}_\mu\partial\phi^B}}\,\eta^B
+{{\partial{\cal L}_0}\over{\partial{\phi^A}_\mu\partial{\phi^B}_\nu}}\,{\eta^B}_\nu\right)\,.
\label{1.11}
\end{eqnarray}

\noindent Equating $J_A(j^2\eta)$ to zero, {\it i.e.} setting

\begin{equation}
{\rm Jac}_{{\cal L}_0}(j^2\eta)=0\,,
\label{1.12}
\end{equation}

\noindent gives rise to the standard form of the {\it (generalised) Jacobi equation}.

Recalling that there is a natural bundle isomorphism ${\cal I}:V(J^1\pi)\rightarrow J^1(V\pi)$, locally defined by:

\begin{equation}
(x^\lambda,\,y^i,\,{y^i}_\mu;\,\eta^i,\,{\eta^i}_\mu)\,\mapsto\,
(x^\lambda,\,y^i,\,\eta^i;\,{y^i}_\mu,\,{\eta^i}_\mu)\,,
\label{1.13}
\end{equation}

\noindent the first variation $\delta{\cal L}_0:V(J^1\pi)\rightarrow\Lambda^n(T^*M)$ of ${\cal L}_0$, locally defined by:

\begin{equation}
\delta{\cal L}_0\equiv{{\partial{\cal L}_0}\over{\partial y^i}}\,\eta^i+{{\partial{\cal L}_0}\over{\partial{y^i}_\mu}}\,{\eta^i}_\mu\,,
\label{1.14}
\end{equation}

\noindent defines in fact a new Lagragian density in the bundle $V\pi$ by the prescription:

\begin{equation}
{\cal L}_1=\delta{\cal L}_0\,\circ\,({\cal I})^{-1}\,.
\label{1.15}
\end{equation}

\noindent The new Lagrangian ${\cal L}_1$ is the {\it first--order deformed Lagrangian}. It is not difficult to prove the following:

\bigskip

\noindent{\bf Theorem 1.} {\it Let $(B,M,\pi)$ be any fibered manifold and ${\cal L}_0$ be any first order lagrangian density on $\pi$. Then the system formed by the Euler--Lagrange equation (\ref{1.3}) and the Jacobi equation (\ref{1.12}) of ${\cal L}_0$ is equivalent to the Euler--Lagrange equations of ${\cal L}_1$.}

\bigskip

\noindent This result, already used in \cite{11} for the purpose of application to Riemannian Geometry, will not be deduced here. It will in fact follow from the analogous statement for second--order Lagrangians which will be discussed in the seubsequant Sections of this paper.

\setcounter{equation}{0}
\section{Field Theory}

We discuss now the generalisation of these results, including the Jacobi equation as developed in \cite{05}, for general second--order field theories. For the sake of immediate understanding, the discussion will be given first in a coordinate language and using the analytical technique of truncated power expansions to express homotopic variations. Intrinsic geometric expressions will also be briefly discussed. The reader is referred to \cite{05} for further details about notation.

\subsection{Series Development of the Action and Euler--Lagrange\\ Equations}

Let us start by considering an action

\begin{equation}
{\cal S}=\int_\Omega\,{\cal L}_0(x,\,\phi,\,\partial\phi,\,\partial^2\phi)\,ds\,,
\label{2.1}
\end{equation}

\noindent and let us consider an infinitesimal expansion of the action ${\cal S}$ around a section ${\phi^A}_0$. The infinitesimal deformations of $\phi^A$ together with its derivatives ${\phi^A}_\mu$ and ${\phi^A}_{\mu\nu}$ are given by

\begin{eqnarray}
\phi^A&=&{\phi^A}_0+\epsilon\,\eta^A+{1\over2}\,\epsilon^2\,\rho^A+o(\epsilon^3)\,,\nonumber\\
{\phi^A}_\mu&=&\left({\phi^A}_0\right)_\mu+\epsilon\,{\eta^A}_\mu+{1\over2}\,\epsilon^2\,{\rho^A}_\mu+o(\epsilon^3)\,,\nonumber\\
{\phi^A}_{\mu\nu}&=&\left({\phi^A}_0\right)_{\mu\nu}+\epsilon\,{\eta^A}_{\mu\nu}+{1\over2}\,
\epsilon^2\,{\rho^A}_{\mu\nu}+o(\epsilon^3)\,,
\label{2.2}
\end{eqnarray}

\noindent where $\epsilon$ is a smallness parameter, while $\eta=\eta^A\partial_A$ and $\rho=\rho^A\partial_A$ are vertical vectorfields which correspond to the ``classical'' first and second variation $\delta\phi$ and $\delta^2\phi$ respectively. The variation of the action, at the second--order in $\epsilon$, is given by

\begin{equation}
{\cal S}(\epsilon)={\cal S}_0+\epsilon\,\delta{\cal S}+{1\over2}\,\epsilon^2\,(\delta^2{\cal S})
+\cdots\,,
\label{2.3}
\end{equation}

\noindent where

\begin{eqnarray}
\delta{\cal S}&=&\int_\Omega\,\left[{{\partial{\cal L}_0}\over{\partial\phi^A}}\,\eta^A
+{{\partial{\cal L}_0}\over{\partial{\phi^A}_\mu}}\,{\eta^A}_\mu
+{{\partial{\cal L}_0}\over{\partial{\phi^A}_{\mu\nu}}}\,{\eta^A}_{\mu\nu}\right]\,ds\,,
\nonumber\\
\delta^2{\cal S}&=&\int_\Omega\,\left[{{\delta{\cal L}_0}\over{\delta\phi^A}}\,\rho^A
+{{\delta{\cal L}_0}\over{\delta{\phi^A}_\mu}}\,{\rho^A}_\mu
+{{\delta{\cal L}_0}\over{\delta{\phi^A}_{\mu\nu}}}\,{\rho^A}_{\mu\nu}\right]\,ds
+\int_\Omega\,{\cal L}_2\,ds\,,
\label{2.4}
\end{eqnarray}

\noindent and

\begin{eqnarray}
{\cal L}_2&=&{{\partial^2{\cal L}_0}\over{\partial\phi^A\partial\phi^B}}\,\eta^A\,\eta^B
+2\,{{\partial^2{\cal L}_0}\over{\partial\phi^A\partial{\phi^B}_\mu}}\,\eta^A\,{\eta^B}_\mu
+2\,{{\partial^2{\cal L}_0}\over{\partial\phi^A\partial{\phi^B}_{\mu\nu}}}\,\eta^A\,
{\eta^B}_{\mu\nu}\nonumber\\
&&+{{\partial^2{\cal L}_0}\over{\partial{\phi^A}_\mu\partial{\phi^B}_\nu}}\,{\eta^A}_\mu\,
{\eta^B}_\nu
+2\,{{\partial^2{\cal L}_0}\over{\partial{\phi^A}_\mu\partial{\phi^B}_{\nu\lambda}}}\,
{\eta^A}_\mu\,{\eta^B}_{\nu\lambda}
+{{\partial^2{\cal L}_0}\over{\partial{\phi^A}_{\mu\nu}\partial{\phi^B}_{\lambda\rho}}}\,
{\eta^A}_{\mu\nu}\,{\eta^B}_{\lambda\rho}\,,\nonumber\\&&
\label{2.5}
\end{eqnarray}

\noindent is the Hessian of ${\cal L}_0$, which by later convenience we have denoted by ${\cal L}_2$.

With the expansion above, we can now deal with the prob;em of extremality and stability of the action. Hereafter, the original Lagrangian ${\cal L}_0$ will be simply denoted by ${\cal L}$ to simplify notation.

Let us then assume that $\phi^A={\phi^A}_0(x)$ is a section along which ${\cal S}$ has an extremum, {\it i.e.} the first variation of the action vanishes for all first--order deformations $\eta^A$. To this purpose it is convenient to perform an integration by parts and rewrite $\delta{\cal S}$ as follows:

\begin{equation}
\delta{\cal S}=\int_\Omega\,{{\delta{\cal L}}\over{\delta\phi^A}}\,\eta^A\,ds
+\oint_{\partial\Omega}\,\left[{{\delta{\cal L}}\over{\delta{\phi^A}_\mu}}\,\eta^A
+{{\delta{\cal L}}\over{\delta{\phi^A}_{\mu\nu}}}\,{\eta^A}_\nu\right]\,ds_\mu\,,
\label{2.6}
\end{equation}

\noindent where

\begin{eqnarray}
{{\delta{\cal L}}\over{\delta\phi^A}}&=&
{{\partial{\cal L}}\over{\partial\phi^A}}-{d\over{dx^\mu}}\left({{\partial{\cal L}}\over{\partial{\phi^A}_\mu}}\right)
+{{d^2}\over{dx^\mu dx^\nu}}\left({{\partial{\cal L}}\over{\partial{\phi^A}_{\mu\nu}}}\right)=
{{\partial{\cal L}}\over{\partial\phi^A}}-{d\over{dx^\mu}}\left({{\delta{\cal L}}\over{\delta{\phi^A}_\mu}}\right)\,,\nonumber\\
{{\delta{\cal L}}\over{\delta{\phi^A}_\mu}}&=&
{{\partial{\cal L}}\over{\partial{\phi^A}_\mu}}-{d\over{dx^\nu}}\left({{\partial{\cal L}}\over{\partial{\phi^A}_{\mu\nu}}}\right)\,,
\label{2.7}
\end{eqnarray}

\noindent are the Euler--Lagrange derivatives.

As in the first--order case (see \cite{11}), the local coordinate expression (\ref{2.6}) corresponds to the intrinsic decomposition

\begin{equation}
\delta{\cal S}=\int_\Omega\,\left<e({\cal L})|\eta\right>+\int_{\partial\Omega}\,\left<f({\cal L})|j^1\eta\right>\,,
\label{2.8}
\end{equation}

\noindent where $e({\cal L}):J^4\pi\rightarrow\Lambda^m(T^*M)\otimes V^*\pi$ is the Euler--Lagrange morphism and $f({\cal L}):J^3\pi\rightarrow\Lambda^{m-1}(T^*M)\otimes V^*\pi^1$ is the canonical Poincar\'e--Cartan morphism (uniquely existing for second--order theories, as is well known from the general theory; see \cite{12}, \cite{13} and references quoted therein).

The extremality of the action is classically described by the Euler--Lagrange equations

\begin{equation}
{{\delta{\cal L}}\over{\delta\phi^A}}=0\,,
\label{2.9}
\end{equation}

\noindent under the boundary conditions

\begin{eqnarray}
\left.\eta^A\right|_{\partial\Omega}&=&0\,,\nonumber\\
\left.{\eta^A}_\mu\right|_{\partial\Omega}&=&0\,,
\label{2.10}
\end{eqnarray}

\noindent which just mean that the deformation of the local section $\phi_0$ is fixed at the boundary, together with its first derivatives. The above conditions just concern the extremality of the action. They do not say anything about the stability of the solutions and to this purpose we must go to the next order of deformations.

\subsection{The Jacobi Equation}

A criterion for the stability of the Euler--Lagrange equations is obtained by looking at the sign of the second variation $\delta^2{\cal S}$. If $\delta^2{\cal S}>0$ we have a minimum; if $\delta^2{\cal S}<0$ we have a maximum; if $\delta^2{\cal S}=0$ we might have a degenerate critical point. Assuming that we are ``on shell'', {\it i.e.} that the Euler--Lagrange equations hold along the section $\phi_0$, then $\delta^2{\cal S}$ reduces to

\begin{equation}
\delta^2{\cal S}=\oint_{\partial\Omega}\,\left({{\delta{\cal L}_0}\over{\delta{\phi^A}_\mu}}\,
\rho^A+{{\delta{\cal L}_0}\over{\delta{\phi^A}_{\mu\nu}}}\,{\rho^A}_\nu\right)\,ds_\mu+\int_
\Omega\,{\cal L}_2\,ds\,.
\label{2.11}
\end{equation}

\noindent In order for $\delta^2{\cal S}$ to have a definite sign we impose the further boundary conditions

\begin{eqnarray}
\left.\rho^A\right|_{\partial\Omega}&=&0\,,\nonumber\\
\left.{\rho^A}_\mu\right|_{\partial\Omega}&=&0\,,
\label{2.12}
\end{eqnarray}

\noindent {\it i.e.}, we require that $\phi_0$ is strongly fixed at the boundary up to order two. Under these stronger conditions $\delta^2{\cal S}$ reduces to

\begin{equation}
\delta^2{\cal S}=\int_\Omega\,{\cal L}_2\,ds\,.
\label{2.13}
\end{equation}

As already discussed in \cite{05} the Jacobi equation can now be obtained by a suitable integration by parts. Let us however remark that the quadratic form appearing in (\ref{2.13}) can be changed by adding surface terms which, due to conditions (\ref{2.10}) vanish at the integration boundary. Therefore, the form of the Jacobi equation to which one arrives strongly depends on the terms added at the boundary. One criteria would be to add surface terms which do not change the dependence on $(\eta,\partial\eta,\partial^2\eta)$. For this let us rewrite ${\cal L}_2$ in the form

\begin{eqnarray}
{\cal L}_2&=&
{{\partial^2{\cal L}}\over{\partial\phi^A\partial\phi^B}}\,\eta^A\,\eta^B
+{{\partial^2{\cal L}}\over{\partial\phi^A\partial{\phi^B}_\mu}}\,\eta^A\,{\eta^B}_\mu
+{{\partial^2{\cal L}}
\over{\partial\phi^A\partial{\phi^B}_{\mu\nu}}}\,\eta^A\,{\eta^B}_{\mu\nu}\nonumber\\
&&+{{\partial^2{\cal L}}\over{\partial{\phi^A}_\mu\partial\phi^B}}\,{\eta^A}_\mu\,\eta^B
+{{\partial^2{\cal L}}\over{\partial{\phi^A}_\mu\partial{\phi^B}_\nu}}\,{\eta^A}_\mu\,
{\eta^B}_\nu
+{{\partial^2{\cal L}}\over{\partial{\phi^A}_\mu\partial{\phi^B}_{\nu\lambda}}}\,
{\eta^A}_\mu\,{\eta^B}_{\nu\lambda}\nonumber\\
&&+{{\partial^2{\cal L}}\over{\partial{\phi^A}_{\mu\nu}\partial\phi^B}}\,{\eta^A}_{\mu\nu}\,
\eta^B
+{{\partial^2{\cal L}}\over{\partial{\phi^A}_{\mu\nu}\partial{\phi^B}_\lambda}}\,
{\eta^A}_{\mu\nu}\,{\eta^B}_\lambda
+{{\partial^2{\cal L}}\over{\partial{\phi^A}_{\mu\nu}
\partial{\phi^B}_{\lambda\rho}}}\,{\eta^A}_{\mu\nu}\,{\eta^B}_{\lambda\rho}\,.\nonumber\\&&
\label{2.14}
\end{eqnarray}

The only terms which can be integrated by parts are the fourth, fifth, seventh and eighth terms. The four possible outcomes are summarised hereafter:

\begin{eqnarray}
{\cal L}_2&\cong&
{\partial\over{\partial\phi^A}}\left({{\partial{\cal L}}\over{\partial\phi^B}}-{d\over{dx^\mu}}
\left({{\partial{\cal L}}\over{\partial{\phi^B}_\mu}}\right)\right)\,\eta^A\,\eta^B
+\left({{\partial^2{\cal L}}\over{\partial\phi^A\partial{\phi^B}_\mu}}-{{\partial^2{\cal L}}
\over{\partial\phi^B\partial{\phi^A}_\mu}}\right)\,\eta^A\,{\eta^B}_\mu\nonumber\\
&&+2\,{{\partial^2{\cal L}}\over{\partial\phi^A\partial{\phi^B}_{\mu\nu}}}\,\eta^A\,{\eta^B}
_{\mu\nu}
+{{\partial^2{\cal L}}\over{\partial{\phi^A}_\mu\partial{\phi^B}_\nu}}\,{\eta^A}_\mu\,{\eta^B}
_\nu
+2\,+{{\partial^2{\cal L}}\over{\partial{\phi^A}_\mu\partial{\phi^B}_{\nu\lambda}}}\,{\eta^A}
_\mu\,{\eta^B}_{\nu\lambda}\nonumber\\
&&+{{\partial^2{\cal L}}\over{\partial{\phi^A}_{\mu\nu}\partial{\phi^B}_{\lambda\rho}}}\,
{\eta^A}_{\mu\nu}\,{\eta^B}_{\lambda\rho}
+{d\over{dx^\mu}}\left({{\partial^2{\cal L}}\over{\partial\phi^A\partial{\phi^B}_\mu}}\,\eta^A
\,\eta^B\right)\,,\nonumber\\
{\cal L}_2&\cong&
{{\partial^2{\cal L}}\over{\partial\phi^A\partial\phi^B}}\,\eta^A\,\eta^B
+\left(2\,{{\partial^2{\cal L}}\over{\partial\phi^A\partial{\phi^B}_\mu}}-{d\over{dx^\nu}}
\left({{\partial^2{\cal L}}\over{\partial{\phi^A}_\mu\partial{\phi^B}_\nu}}\right)\right)\,
\eta^A\,{\eta^B}_\mu\nonumber\\
&&+\left(2\,{{\partial^2{\cal L}}\over{\partial\phi^A\partial{\phi^B}_{\mu\nu}}}-{{\partial^2{\cal L}}\over{\partial{\phi^A}_\mu\partial{\phi^B}_\nu}}\right)\,
\eta^A\,{\eta^B}_{\mu\nu}
+2\,{{\partial^2{\cal L}}\over{\partial{\phi^A}_\mu\partial{\phi^B}_{\nu\lambda}}}\,
{\eta^A}_\mu\,{\eta^B}_{\nu\lambda}\nonumber\\
&&+{{\partial^2{\cal L}}\over{\partial{\phi^A}_{\mu\nu}\partial{\phi^B}_{\lambda\rho}}}\,
{\eta^A}_{\mu\nu}\,{\eta^B}_{\lambda\rho}
+{d\over{dx^\mu}}\left({{\partial^2{\cal L}}\over{\partial{\phi^A}_\mu\partial{\phi^B}_\nu}}\,
\eta^A\,{\eta^B}_\nu\right)\,,\nonumber\\
{\cal L}_2&\cong&
{{\partial^2{\cal L}}\over{\partial\phi^A\partial\phi^B}}\,\eta^A\,\eta^B
+\left(2\,{{\partial^2{\cal L}}\over{\partial\phi^A\partial{\phi^B}_\mu}}-{\partial\over{
\partial\phi^A}}{d\over{dx^\nu}}\left({{\partial{\cal L}}\over{\partial{\phi^B}_{\mu\nu}}}
\right)\right)\,\eta^A\,{\eta^B}_\mu\nonumber\\
&&+{{\partial^2{\cal L}}\over{\partial\phi^A\partial{\phi^B}_{\mu\nu}}}\,\eta^A\,
{\eta^B}_{\mu\nu}
+\left({{\partial^2{\cal L}}\over{\partial{\phi^A}_\mu\partial{\phi^B}_\nu}}-{{\partial^2{\cal L}}\over{\partial\phi^A\partial{\phi^B}_{\mu\nu}}}\right)\,{\eta^A}_\mu\,{\eta^B}_\mu\nonumber\\
&&+2\,{{\partial^2{\cal L}}\over{\partial{\phi^A}_\mu\partial{\phi^B}_{\nu\lambda}}}\,
{\eta^A}_\mu\,{\eta^B}_{\nu\lambda}
+{{\partial^2{\cal L}}\over{\partial{\phi^A}_{\mu\nu}\partial{\phi^B}_{\lambda\rho}}}
\,{\eta^A}_{\mu\nu}\,{\eta^B}_{\lambda\rho}\nonumber\\
&&+{d\over{dx^\mu}}\left({{\partial^2{\cal L}}\over{\partial\phi^A\partial{\phi^B}_{\mu\nu}}}\,\eta^A\,{\eta^B}_\nu\right)\,,\nonumber\\
{\cal L}_2&\cong&
{{\partial^2{\cal L}}\over{\partial\phi^A\partial\phi^B}}\,\eta^A\,\eta^B
+2\,{{\partial^2{\cal L}}\over{\partial\phi^A\partial{\phi^B}_\mu}}\,\eta^A\,{\eta^B}_\mu
+2\,{{\partial^2{\cal L}}\over{\partial\phi^A\partial{\phi^B}_{\mu\nu}}}\,\eta^A\,
{\eta^B}_{\mu\nu}\nonumber\\
&&+\left({{\partial^2{\cal L}}\over{\partial{\phi^A}_\mu\partial{\phi^B}_\nu}}-{d\over{dx^
\lambda}}\left({{\partial^2{\cal L}}\over{\partial{\phi^A}_\mu\partial{\phi^B}_{\nu\lambda}}}
\right)\right)\,{\eta^A}_\mu\,{\eta^B}_\nu\nonumber\\
&&+\left({{\partial^2{\cal L}}\over{\partial{\phi^A}_\mu\partial{\phi^B}_{\nu\lambda}}}
-{{\partial^2{\cal L}}\over{\partial{\phi^B}_\mu\partial{\phi^A}_{\nu\lambda}}}\right)\,
{\eta^A}_\mu\,{\eta^B}_{\nu\lambda}
+{{\partial^2{\cal L}}\over{\partial{\phi^A}_{\mu\nu}\partial{\phi^B}_{\lambda\rho}}}\,
{\eta^A}_{\mu\nu}\,{\eta^B}_{\lambda\rho}\nonumber\\
&&+{d\over{dx^\mu}}\left({{\partial^2{\cal L}}\over{\partial{\phi^A}_{\mu\nu}\partial{\phi^B}_
\lambda}}\,{\eta^A}_\nu\,{\eta^B}_\lambda\right)\,,
\label{2.15}
\end{eqnarray}

\noindent where $\cong$ means equal modulo a total divergence.

Apart from first--order divergencies, we can also add second--order divergencies to the Hessian. The only two possibilities which do not introduce derivatives higher than fourth--order ones are the following

\begin{eqnarray}
&&{{d^2}\over{dx^\mu dx^\nu}}\left({{\partial^2{\cal L}}\over{\partial{\phi^A}_\mu\partial
{\phi^B}_\nu}}\,\eta^A\,\eta^B\right)\nonumber\\
&=&{{d^2}\over{dx^\mu dx^\nu}}\left({{\partial^2{\cal L}}\over{\partial{\phi^A}_\mu\partial{
\phi^B}_\nu}}\right)\,\eta^A\,\eta^B
+4\,{d\over{dx^\mu}}\left({{\partial^2{\cal L}}\over{\partial{\phi^A}_\mu\partial{\phi^B}_\nu}}
\right)\,\eta^A\,{\eta^B}_\nu\nonumber\\
&&+2\,{{\partial^2{\cal L}}\over{\partial{\phi^A}_\mu\partial{\phi^B}_\nu}}\,\eta^A\,{\eta^B}
_{\mu\nu}
+2\,{{\partial^2{\cal L}}\over{\partial{\phi^A}_\mu\partial{\phi^B}_\nu}}\,{\eta^A}_\mu\,
{\eta^B}_\nu\,,\nonumber\\
&&{{d^2}\over{dx^\mu dx^\nu}}\left({{\partial^2{\cal L}}\over{\partial\phi^A\partial{\phi^B}
_{\mu\nu}}}\,\eta^A\,\eta^B\right)\nonumber\\
&=&{{d^2}\over{dx^\mu dx^\nu}}\left({{\partial^2{\cal L}}\over{\partial\phi^A\partial{\phi^B}_
{\mu\nu}}}\right)\,\eta^A\,\eta^B
+2\,{d\over{dx^\mu}}\left({{\partial^2{\cal L}}\over{\partial\phi^A\partial{\phi^B}_{\mu\nu}}}
+{{\partial^2{\cal L}}\over{\partial\phi^B\partial{\phi^A}_{\mu\nu}}}\right)\,\eta^A\,
{\eta^B}_\nu\nonumber\\
&&+\left({{\partial^2{\cal L}}\over{\partial\phi^A\partial{\phi^B}_{\mu\nu}}}+{{\partial^2{\cal L}}\over{\partial\phi^B\partial{\phi^A}_{\mu\nu}}}\right)\,\eta^A\,{\eta^B}_{\mu\nu}
+2\,{{\partial^2{\cal L}}\over{\partial\phi^A\partial{\phi^B}_{\mu\nu}}}\,{\eta^A}_\mu\,
{\eta^B}_\nu\,.
\label{2.16}
\end{eqnarray}

A simpler possibility is to combine all the above integration by parts in a single decomposition displaying better properties. The result similar to the analogous one for first--order theories (see \cite{05}) is given by

\begin{eqnarray}
{\cal L}_2&\cong&{\partial\over{\partial\phi^A}}\left({{\delta{\cal L}}\over{\delta\phi^B}}
\right)\,\eta^A\,\eta^B
+\left({{\partial^2{\cal L}}\over{\partial\phi^A\partial{\phi^B}_\mu}}-{{\partial^2{\cal L}}\over{\partial\phi^B\partial{\phi^A}_\mu}}\right)\,\eta^A\,{\eta^B}_\mu\nonumber\\
&&+\left({{\partial^2{\cal L}}\over{\partial{\phi^A}_\mu\partial{\phi^B}_\nu}}-{{\partial^2{\cal L}}\over{\partial\phi^A\partial{\phi^B}_{\mu\nu}}}\right)\,{\eta^A}_\mu\,{\eta^B}_\nu\nonumber\\
&&+\left({{\partial^2{\cal L}}\over{\partial{\phi^A}_\mu\partial{\phi^B}_{\nu\lambda}}}-{{\partial^2{\cal L}}\over{\partial{\phi^B}_\mu\partial{\phi^A}_{\nu\lambda}}}\right)\,
{\eta^A}_\mu\,{\eta^B}_{\nu\lambda}
+{{\partial^2{\cal L}}\over{\partial{\phi^A}_{\mu\nu}\partial{\phi^B}_{\lambda\rho}}}\,
{\eta^A}_{\mu\nu}\,{\eta^B}_{\lambda\rho}\nonumber\\
&&+{d\over{dx^\mu}}\left({\partial\over{\partial\phi^A}}\left({{\delta{\cal L}}\over{\delta
{\phi^B}_\mu}}\right)\,\eta^A\,\eta^B
+{{\partial^2{\cal L}}\over{\partial\phi^A\partial{\phi^B}_{\mu\nu}}}\,\eta^A\,{\eta^B}_\nu
+{{\partial^2{\cal L}}\over{\partial{\phi^A}_\mu\partial{\phi^B}_{\nu\lambda}}}\,{\eta^A}_\nu
\,{\eta^B}_\lambda\right)\,.\nonumber\\
&&
\label{2.17}
\end{eqnarray}

However, other terms can be added to the Hessian changing the dependence on only $(\eta,\partial\eta,\partial^2\eta)$. An interesting one is the following further expression:

\begin{eqnarray}
{\cal L}_2&\cong&\left[{{\partial^2{\cal L}}\over{\partial\phi^A\partial\phi^B}}\,\eta^B
+{{\partial^2{\cal L}}\over{\partial\phi^A\partial{\phi^B}_\lambda}}\,{\eta^B}_\lambda
+{{\partial^2{\cal L}}\over{\partial\phi^A\partial{\phi^B}_{\lambda\rho}}}\,
{\eta^B}_{\lambda\rho}\right.\nonumber\\
&&-{d\over{dx^\mu}}\left({{\partial^2{\cal L}}\over{\partial{\phi^A}_\mu\partial\phi^B}}\,\eta^B
+{{\partial^2{\cal L}}\over{\partial{\phi^A}_\mu\partial{\phi^B}_\lambda}}\,{\eta^B}_\lambda
+{{\partial^2{\cal L}}\over{\partial{\phi^A}_\mu\partial{\phi^B}_{\lambda\rho}}}\,
{\eta^B}_{\lambda\rho}\right)\nonumber\\
&&\left.+{{d^2}\over{dx^\mu dx^\nu}}\left({{\partial^2{\cal L}}\over{\partial{\phi^A}_{\mu\nu}
\partial\phi^B}}\,\eta^B+{{\partial^2{\cal L}}\over{\partial{\phi^A}_{\mu\nu}\partial
{\phi^B}_\lambda}}\,{\eta^B}_\lambda+{{\partial^2{\cal L}}\over{\partial{\phi^A}_{\mu\nu}
\partial{\phi^B}_{\lambda\rho}}}\,{\eta^B}_{\lambda\rho}\right)\right]\,\eta^A\nonumber\\
&&-{d\over{dx^\mu}}\left[\left({{\partial^2{\cal L}}\over{\partial{\phi^A}_\mu\partial\phi^B}}
\,\eta^B+{{\partial^2{\cal L}}\over{\partial{\phi^A}_\mu\partial{\phi^B}_\lambda}}\,{\eta^B}_
\lambda+{{\partial^2{\cal L}}\over{\partial{\phi^A}_\mu\partial{\phi^B}_{\lambda\rho}}}\,
{\eta^B}_{\lambda\rho}\right.\right.\nonumber\\
&&\left.-{d\over{dx^\nu}}\left({{\partial^2{\cal L}}\over{\partial{\phi^A}_{\mu\nu}\partial
\phi^B}}\,\eta^B+{{\partial^2{\cal L}}\over{\partial{\phi^A}_{\mu\nu}\partial{\phi^B}_\lambda}}
\,{\eta^B}_\lambda+{{\partial^2{\cal L}}\over{\partial{\phi^A}_{\mu\nu}\partial{\phi^B}_{\lambda
\rho}}}\,{\eta^B}_{\lambda\rho}\right)\right)\,\eta^A\nonumber\\
&&\left.+\left({{\partial^2{\cal L}}\over{\partial{\phi^A}_{\mu\nu}\partial\phi^B}}\,\eta^B
+{{\partial^2{\cal L}}\over{\partial{\phi^A}_{\mu\nu}\partial{\phi^B}_\lambda}}\,{\eta^B}
_\lambda+{{\partial^2{\cal L}}\over{\partial{\phi^A}_{\mu\nu}\partial{\phi^B}_{\lambda\rho}}}
\,{\eta^B}_{\lambda\rho}\right)\,{\eta^A}_\nu\right]\,.
\label{2.18}
\end{eqnarray}

In all the above cases the surface terms cancel at the integration boundary. therefore, different Jacobi equations would be obtained. The equation ensuing from (\ref{2.18}), namely

\begin{eqnarray}
{{\partial^2{\cal L}}\over{\partial\phi^A\partial\phi^B}}\,\eta^B+{{\partial^2{\cal L}}\over{
\partial\phi^A\partial{\phi^B}_\lambda}}\,{\eta^B}_\lambda+{{\partial^2{\cal L}}\over{\partial
\phi^A\partial{\phi^B}_{\lambda\rho}}}\,{\eta^B}_{\lambda\rho}&&\nonumber\\
-{d\over{dx^\mu}}\left({{\partial^2{\cal L}}\over{\partial{\phi^A}_\mu\partial\phi^B}}\,\eta^B
+{{\partial^2{\cal L}}\over{\partial{\phi^A}_\mu\partial{\phi^B}_\lambda}}\,{\eta^B}_\lambda
+{{\partial^2{\cal L}}\over{\partial{\phi^A}_\mu\partial{\phi^B}_{\lambda\rho}}}\,{\eta^B}_{
\lambda\rho}\right)&&\nonumber\\
+{{d^2}\over{dx^\mu dx^\nu}}\left({{\partial^2{\cal L}}\over{\partial{\phi^A}_{\mu\nu}\partial
\phi^B}}\,\eta^B+{{\partial^2{\cal L}}\over{\partial{\phi^A}_{\mu\nu}\partial{\phi^B}_\lambda
}}\,{\eta^B}_\lambda+{{\partial^2{\cal L}}\over{\partial{\phi^A}_{\mu\nu}\partial{\phi^B}_{
\lambda\rho}}}\,{\eta^B}_{\lambda\rho}\right)&=&0\,,
\label{2.19}
\end{eqnarray}

\noindent is the ``standard Jacobi equation'' for ${\cal L}$. It is the equation (\ref{0.2}) we already mentioned in the Introduction

\newpage
\setcounter{equation}{0}
\section{Hierarchical Structure of the Deformed Lagrangian}

Now we introduce a hierarchical structure associated to the deformations of the Lagrangian.

\subsection{The Deformed Lagrangian}

Let us again consider an infinitesimal deformation of the field $\phi^A$ and its derivatives ${\phi^A}_\mu$ and ${\phi^A}_{\mu\nu}$, up to second--oder in $\epsilon$. Then, the variation of the Lagrangian, at second order--order in $\epsilon$, is given by

\begin{equation}
{\cal L}(\eta)={\cal L}_0+\epsilon\,{\cal L}_1(\eta)+{1\over2}\,\epsilon^2\,\left[{\cal L}_2
(\eta)+{\cal L}_1(\rho)\right]\,,
\label{3.1}
\end{equation}

\noindent where ${\cal L}_1$ is given by

\begin{equation}
{\cal L}_1(\eta)={{\partial{\cal L}_0}\over{\partial\phi^A}}\,\eta^A+{{\partial{\cal L}_0}\over
{\partial{\phi^A}_\mu}}\,{\eta^A}_\mu+{{\partial{\cal L}_0}\over{\partial{\phi^A}_{\mu\nu}}}\,
{\eta^A}_{\mu\nu}\,,
\label{3.2}
\end{equation}

\noindent while ${\cal L}_2$ is given by equation (\ref{2.5}).

The introduction of the above definitions is not only a matter of notational convenience since ${\cal L}_0$, ${\cal L}_1$ and ${\cal L}_2$ satisfy a series of remarkable identities. In fact, it can be checked that the Euler--Lagrange derivatives of ${\cal L}_1$ and ${\cal L}_2$ are given by

\begin{eqnarray}
{{\delta{\cal L}_1}\over{\delta\eta^A}}&=&{{\delta{\cal L}_0}\over{\delta\phi^A}}\,,\nonumber\\
{{\delta{\cal L}_2}\over{\delta\eta^A}}&=&{{\delta{\cal L}_1}\over{\delta\phi^A}}\,.
\label{3.3}
\end{eqnarray}

\noindent Moreover, it can be checked (see also \cite{04}) that the equation

\begin{equation}
{{\delta{\cal L}_2}\over{\delta\eta^A}}=0\,,
\label{3.4}
\end{equation}

\noindent corresponds to the Jacobi equation (\ref{2.19}) in its standard form.

\subsection{Unified description of the Euler--Lagrange and Jacobi equations}

Let us now remark that in virtue of (\ref{3.3}) the Euler--Lagrange equation and the Jacobi equation can be rewritten together as

\begin{eqnarray}
{{\delta{\cal L}_0}\over{\delta\phi^A}}&=&{{\delta{\cal L}_1}\over{\delta\eta^A}}=0\,,
\nonumber\\
{{\delta{\cal L}_2}\over{\delta\eta^A}}&=&{{\delta{\cal L}_1}\over{\delta\phi^A}}=0\,.
\label{3.5}
\end{eqnarray}

\noindent Therefore, we can consider the above system of equations as coming from a single variational principle based on the deformed Lagrangian ${\cal L}_1$ where, however, the dynamical variables have been ``doubled'' to $\phi$'s and $\eta$'s.

We can then introduce momenta canonically conjugated to $\phi$'s and $\eta$'s by means of standard prescriptions

\begin{eqnarray}
{\pi_A}^\mu&=&{{\partial{\cal L}_1}\over{\partial{\phi^A}_\mu}}-{d\over{dx^\nu}}
\left({{\partial{\cal L}_1}\over{\partial{\phi^A}_{\mu\nu}}}\right)\,,\nonumber\\
{N_A}^{\mu\nu}&=&{{\partial{\cal L}_1}\over{\partial{\phi^A}_{\mu\nu}}}\,,\nonumber\\
{p_A}^\mu&=&{{\partial{\cal L}_1}\over{\partial{\eta^A}_\mu}}-{d\over{dx^\nu}}
\left({{\partial{\cal L}_1}\over{\partial{\eta^A}_{\mu\nu}}}\right)\,,\nonumber\\
{n_A}^{\mu\nu}&=&{{\partial{\cal L}_1}\over{\partial{\eta^A}_{\mu\nu}}}\,,
\label{3.6}
\end{eqnarray}

\noindent and a canonical energy--momentum tensor by setting:

\begin{equation}
{{\cal H}^\mu}_\nu={\pi_A}^\mu\,{\phi^A}_\nu+{N_A}^{\mu\lambda}\,{\phi^A}_{\lambda\nu}
+{p_A}^\mu\,{\eta^A}_\nu+{n_A}^{\mu\lambda}\,{\eta^A}_{\lambda\nu}-\delta^\mu_\nu\,{\cal L}_1\,.
\label{3.7}
\end{equation}

The explicit expression of ${{\cal H}^\mu}_\nu$ for our case is nor relevant here. What is really important is the fact that, if ${\cal L}_0$ does no depend on ``space--time variables'' $x^\lambda$, then ${{\cal H}^\mu}_\nu$ is a conserved quantity. Therefore, the allowed deformations are selected by the first--order condition

\begin{equation}
{{d{{\cal H}^\mu}_\nu}\over{dx^\mu}}=0\,,
\label{3.8}
\end{equation}

\noindent which can thence replace the second--order condition given by the Jacobi equation (\ref{2.19}).

Let us also remark that the Hessian matrix associated to ${\cal L}_1$ is given by

\begin{equation}
W({\cal L}_1)=\left(\matrix{{{\partial^2{\cal L}_1}\over{\partial{\phi^A}_\mu\partial{\phi^B}_
\nu}}&{{\partial^2{\cal L}_1}\over{\partial{\phi^A}_\mu\partial{\eta^B}_\nu}}\cr&\cr{{\partial^2
{\cal L}_1}\over{\partial{\eta^A}_\mu\partial{\phi^B}_\nu}}&{{\partial^2{\cal L}_1}\over{
\partial{\eta^A}_\mu\partial{\eta^B}_\nu}}\cr}\right)=\left(\matrix{{{\partial^2{\cal L}_1}\over
{\partial{\phi^A}_\mu\partial{\phi^B}_\nu}}&W^{\mu\nu}_{AB}({\cal L}_0)\cr&\cr W^{\mu\nu}_{AB}
({\cal L}_0)&0\cr}\right)\,,
\label{3.9}
\end{equation}

\noindent so that the following holds true

\begin{equation}
\left|\det[W({\cal L}_1)]\right|=[\det[W({\cal L}_0)]]^2\,,
\label{3.10}
\end{equation}

\noindent which ensures that the regularity of ${\cal L}_1$ depends only on the regularity of ${\cal L}_0$.

Suitable globalisations of this concept can be achieved by introducing a connection (see, {\it e.g.}, \cite{14}, \cite{15}). See also \cite{16} and \cite{17} for a discussion of ``$d$--invariance'' of these energy--momentum tensors.

\newpage
\setcounter{equation}{0}
\section{Examples and Applications}

Here we consider some simple Lagrangians which involve up to second--order derivatives in order to illustrate our general results.

\subsection{The Riewe Lagrangian}

The Riewe Lagrangian (see \cite{18}) was introduced to describe a classically spinning particle. It is given by

\begin{equation}
L={1\over2}\,m\,{\dot{\bf q}}^2-{1\over2}\,{m\over{\omega^2}}\,{\ddot{\bf q}}^2\,,
\label{4.1}
\end{equation}

\noindent where ${\bf q}$ is a three--dimensional vector. The equtions of motion are given by

\begin{equation}
{{\delta L}\over{\delta q^i}}=-,\,\delta_{ij}\,\left({\ddot q}^j+{1\over{\omega^2}}\,{\ddot{
\ddot q}}^j\right)=0\,.
\label{4.2}
\end{equation}

\noindent The solutions of the above equation of motion are given by

\begin{equation}
q^i=q^i_0+v^i_0\,t+a^i\,\cos(\omega\,t)+b^i\,\sin(\omega\,t)\,.
\label{4.3}
\end{equation}

\noindent These solutions describe a helical motion along a ellipsoid of semiaxis $a^i$ and $b^i$ oriented in the direction of $v^i$.

The corresponding Jacobi equation is given by

\begin{equation}
{\ddot\eta}^i+{1\over{\omega^2}}\,{\ddot{\ddot\eta}}^i=0.
\label{4.4}
\end{equation}

\subsection{The Shadwick Lagrangian}

The Shadwick Lagrangian is given by

\begin{equation}
{\cal L}=\phi\,\left(\phi_{00}\,\phi_{11}-\phi_{01}^2\right)\,,
\label{4.5}
\end{equation}

\noindent where $\phi$ is a scalar field (see \cite{19}). The field equations are given by

\begin{equation}
{{\delta{\cal L}}\over{\delta\phi}}=3\,\left(\phi_{00}\,\phi_{11}-\phi_{01}^2\right)=0\,.
\label{4.6}
\end{equation}

\noindent The corresponding Jacobi equation is given by

\begin{equation}
\phi_{11}\,\eta_{00}+\phi_{00}\,\eta_{11}-2\,\phi_{01}\,\eta_{01}=0\,.
\label{4.7}
\end{equation}

\subsection{The Klein--Gordon equation}

The Klein--Gordon equation is obtained from the Lagrangian density

\begin{equation}
{\cal L}_0={1\over2}\,\left(g^{\mu\nu}\,\phi_\mu\,\phi_\nu-m^2\,\phi^2\right)\,,
\label{4.8}
\end{equation}

\noindent whose first--order and second--order deformed Lagrangians are given by

\begin{eqnarray}
{\cal L}_1&=&g^{\mu\nu}\,\phi_\mu\,\eta_\nu-m^2\,\phi\,\eta\,,\nonumber\\
{\cal L}_2&=&{1\over2}\,\left(g^{\mu\nu}\,\eta_\mu\,\eta_\nu-m^2\,\eta^2\right)\,.
\label{4.9}
\end{eqnarray}

\noindent The Euler--Lagrange and Jacobi equations are given by

\begin{eqnarray}
{{\delta{\cal L}_1}\over{\delta\eta}}&=&-\left(g^{\mu\nu}\,\phi_{\mu\nu}+m^2\,\phi\right)=0\,,
\nonumber\\
{{\delta{\cal L}_1}\over{\delta\phi}}&=&-\left(g^{\mu\nu}\,\eta_{\mu\nu}+m^2\,\eta\right)=0\,,
\label{4.10}
\end{eqnarray}

\noindent respectively. Solutions to these equations are expressed by

\begin{eqnarray}
\phi(x,\,t)&=&\int\,\left(f_+(k)\,e^{i(k\,x+\sqrt{k^2+m^2}\,t)}
+f_-(k)\,e^{i(k\,x-\sqrt{k^2+m^2}\,t)}\right)\,dk\,,\nonumber\\
\eta(x,\,t)&=&\int\,\left(h_+(k)\,e^{i(k\,x+\sqrt{k^2+m^2}\,t)}
+h_-(k)\,e^{i(k\,x-\sqrt{k^2+m^2}\,t)}\right)\,dk\,,
\label{4.11}
\end{eqnarray}

\noindent such that

\begin{eqnarray}
\phi_0(x,\,t)&=&i\,\int\,\sqrt{k^2+m^2}\,\left(f_+(k)\,e^{i(k\,x+\sqrt{k^2+m^2}\,t)}
-f_-(k)\,e^{i(k\,x-\sqrt{k^2+m^2}\,t)}\right)\,dk\,,\nonumber\\
\phi_i(x,\,t)&=&i\,\int\,k_i\,\left(f_+(k)\,e^{i(k\,x+\sqrt{k^2+m^2}\,t)}
+f_-(k)\,e^{i(k\,x-\sqrt{k^2+m^2}\,t)}\right)\,dk\,,\nonumber\\
\eta_0(x,\,t)&=&i\,\int\,\sqrt{k^2+m^2}\,\left(h_+(k)\,e^{i(k\,x+\sqrt{k^2+m^2}\,t)}
-h_-(k)\,e^{i(k\,x-\sqrt{k^2+m^2}\,t)}\right)\,dk\,,\nonumber\\
\eta_i(x,\,t)&=&i\,\int\,k_i\,\left(h_+(k)\,e^{i(k\,x+\sqrt{k^2+m^2}\,t)}
+h_-(k)\,e^{i(k\,x-\sqrt{k^2+m^2}\,t)}\right)\,dk\,.
\label{4.12}
\end{eqnarray}

\noindent The appropriate energy--momentum tensor is thence given by

\begin{equation}
{{\cal H}^\mu}_\nu({\cal L}_1)=\phi_\nu\,g^{\mu\lambda}\,\phi_\lambda+\eta_\nu\,g^{\mu\lambda}\,
\eta_\lambda-\delta^\mu_\nu\,{\cal L}_1\,,
\label{4.13}
\end{equation}

\noindent and its time--time component ${{\cal H}^0}_0({\cal L}_1)$ is given by

\begin{equation}
{{\cal H}^0}_0({\cal L}_1)=\phi_0\,\eta_0+\phi_1\,\eta_1+m^2\,\phi\,\eta\,.
\label{4.14}
\end{equation}

\noindent The associated energy (see also \cite{06},\cite{07}) is therefore given by

\begin{equation}
E({\cal L}_1)=\int\,{{\cal H}^0}_0({\cal L}_1)\,dx\,,
\label{4.15}
\end{equation}

\noindent and an evaluation on--shell gives immediately

\begin{equation}
E({\cal L}_1)=2\,(2\,\pi)^3\,m^2\,\int\,\left(f_+(k)\,h_-(k)+f_-(k)\,h_+(k)\right)\,dk={\rm constant}\,.
\label{4.16}
\end{equation}

\noindent This condition is equivalent to the Jacobi equation.

\subsection{Geodesics in a Riemannian manifold}

A further example of application to standard Jacobi fields along geodesics of a Riemannian manifold $(M,g)$ has been discussed in \cite{11}. Here we can shortly summarise the results.

Let $Q$ be a $n$--dimensional manifold and $(TQ,Q,\tau_Q)$ its tangent bundle. Let $g$ be a Riemannian metric on $Q$. The {\it geodesics} of $(Q,g)$ are those cuves $\gamma:{\bf R}\rightarrow Q$ whose tangent vector ${\dot\gamma}$ is parallel along $\gamma$, {\it i.e.}, it satisfies $\nabla_{{\dot\gamma}}{\dot\gamma}=0$; in local components

\begin{equation}
{\ddot q}^i+{\Gamma^i}_{jk}\,{\dot q}^j\,{\dot q}^k=0\,.
\label{4.17}
\end{equation}

The {\it Jacobi fields} of $(M,g)$ are those vectorfields $\eta=\eta^i\partial_i$, defined along geodesics $\gamma$ by the differential equation:

\begin{equation}
\nabla^2_\gamma\eta+{\rm Riem}(\eta,\,\gamma,\,{\dot\gamma})=0\,,
\label{4.18}
\end{equation}

\noindent where $\nabla^2_\gamma$ denotes the second--order covariant derivative along the curve $\gamma$ and ${\rm Riem}$ is the tri--linear mapping defining the Riemannian curvature of $g$. Jacobi fields generically define infinitesimal deformations of geodesics into families of nearby geodesics. According to \cite{20}, the metric $g$ can be lifted to a metric $g^C$ on the manifold $TQ$, the ``complete lift'', as follows. Let $g=g_{ij}dq^i dq^j$ in a local chart $(U;q^i)$; then the corresponding local expression of $g^C$ in $(TU;q^i,u^i)$ is

\begin{equation}
g^C=2\,g_{ij}\,\delta u^i\,dq^j\,,
\label{4.19}
\end{equation}

\noindent where $\delta u^i$ denotes the following

\begin{equation}
\delta u^i=du^i+{\Gamma^i}_{mk}\,u^m\,dq^k\,.
\label{4.20}
\end{equation}

For any function $f:Q\rightarrow{\bf R}$ a new function $\partial f:TQ\rightarrow{\bf R}$ is defined by setting

\begin{equation}
(\partial f)(q^i,\,u^i)\equiv(\partial_j f)\,u^j\,.
\label{4.21}
\end{equation}

With this notation $g^C$ can be locally expressed by

\begin{equation}
g^C=(\partial g_{ij})\,du^i\,du^j+2\,g_{ij}\,du^i\,dq^j\,,
\label{4.22}
\end{equation}

\noindent {\it i.e.}, the $(2n\times2n)$ matrix of $g^C$ is

\begin{equation}
g^C=\left(\matrix{\partial g_{ij}&g_{ij}\cr g_{ij}&0\cr}\right)\,.
\label{4.23}
\end{equation}

The following holds true:

\bigskip

{\bf Theorem 2.} {\it Let $(Q,g)$ be a Riemannian manifold. Then the system formed by the geodesic equation of $g$ in $Q$ and the Jacobi equation associated to $g$ in $TQ$ is the geodesic equation in $TQ$ of the complete lift metric $g^C$. Therefore this system follows from a variational principle on $TQ$ based on the energy functional defined by the lifted metric $g^C$.}

\bigskip

\noindent We see immediately that this theorem is nothing but a simple consequence of our general results. In fact, the energy functional of $g$ is based on the Lagrangian:

\begin{equation}
{\cal L}={1\over2}\,g_{ij}\,u^i\,u^j\,,
\label{4.24}
\end{equation}

\noindent whose associated first--order deformation Lagrangian is thence given by

\begin{equation}
{\cal L}_1={1\over2}\,(\partial_k g_{ij})\,u^i\,u^j\,\eta^k+g_{ij}\,u^i\,{\dot\eta}^j\,.
\label{4.25}
\end{equation}

\noindent Using $\nabla(g)=0$ this becomes immediately

\begin{equation}
{\cal L}_1=g_{ij}\,\left[{\dot\eta}^i+{\Gamma^i}_{mk}\,u^m\,\eta^k\right]\,u^j\,.
\label{4.26}
\end{equation}

\noindent Then ${\cal L}_1$ is in fact the energy Lagrangian of the lifted metric $g^C=2\,g_{ij}\delta u^i dq^j$.

\section{Conclusions}

We have presented here a natural and direct generalisation of the Jacobi equation to the case of second--order Lagrangians, which are important not only for the sake of completeness but also and specially in view of applications to relativistic field theories (whereby gravitational Lagrangians depend effectively of second--order derivatives of a metric). Concrete applications to relativistic field theories will form the subject of forthcoming investigations.

\section*{Acknowledgements}

One of us (B. C.) acknowledges support of GNSAGA--CNR and of MURST (Nat. Res. Proj. ``Geometria delle Variet\`a Differenziabili'').

One os us (M. F.) acknowledges support of GNFM--CNR and of MURST (Nat. Res. Proj. ``Metodi Geometrici e Probabilistici in Fisica Matematica'').

One of us (V. T.) has been partially supported by a Visiting Professorship of the Italian GNFM--CNR and of the International Centre for Theoretical Physics, Trieste.

\newpage

\end{document}